\documentclass[prd,aps,nofootinbib,superscriptaddress]{revtex4-2}
\usepackage{mathrsfs}
\usepackage{amsfonts}
\usepackage{amsmath}
\usepackage{amssymb}
\usepackage{array}
\usepackage{verbatim}
\usepackage{graphicx}
\usepackage{amsbsy}
\usepackage{bm}
\usepackage[dvipsnames]{xcolor}

\usepackage{slashed}
\usepackage{hyperref}

\begin{document}

\title{Exact Amplitude Reconstruction in the Real Common-Phase Sector of Small-x Diffractive Energy Flow}

\author{Sanskriti Agrawal}
\affiliation{Department of Physics, Aligarh Muslim University, Aligarh - $202001$, India.}

\author{Raktim Abir}
\affiliation{Department of Physics, Aligarh Muslim University, Aligarh - $202001$, India.}
\email{raktim.ph@amu.ac.in}

\begin{abstract}
We develop an operator-level framework for reconstructing the angular structure of small-$x$ diffractive amplitudes from energy-flow measurements. Focusing on coherent diffractive dijet production in the leading-eikonal approximation, we establish the relation between the angular multipoles of the dipole amplitude, the gluon Wigner distribution, and the diffractive scattering amplitude, while keeping their distinct radial transforms explicit. We show that, in the real common-phase sector and away from diffractive zeros, the normalized scattering amplitude can be reconstructed directly, up to a global sign convention, from the square root of the normalized energy-flow distribution. For generic complex amplitudes, the measured intensity instead determines autocorrelations of the amplitude spectrum and phase retrieval is not unique without additional information. We further discuss corrections arising from harmonic-dependent phases, practical conditions for an EIC extraction, and effects beyond the leading-eikonal approximation.

\end{abstract}

\maketitle
%\section{Introduction}
\section{Introduction}
One of the central goals at the Electron-Ion Collider (EIC) is imaging the multidimensional structure of a proton or, more generally, hadrons. While the parton distribution functions (or PDFs) are used to describe the longitudinal momentum carried by the partons, transverse momentum distributions (or TMDs) and generalized parton distributions (or GPDs) provide complementary information about their transverse momentum and spatial distributions, respectively. A more complete and detailed description of partons is provided by generalized transverse-momentum-dependent distributions (or GTMDs) and, after Fourier transformation with respect to the transverse momentum transfer, by Wigner distributions. These distributions encode simultaneous information about the transverse momentum and transverse position of the partons inside hadrons. Complementary light-front studies have constructed gluon GTMDs and Wigner distributions for different proton polarization states, including a recent calculation at nonzero skewness \cite{More:2018glw,Chakrabarti:2025gtmd}. \\

It has been established that diffractive dijet production can probe the orientation of the color dipole amplitude and consequently the correlation between the impact parameter and the dipole size of the $q{\bar q}$ pair \cite{Altinoluk:2015dpi}. In the small-$x$ limit, the Wigner distributions simplify and can be written as Fourier transforms of the well-known impact-parameter-dependent dipole amplitudes \cite{Hatta:2016dxp}. This relation remains useful in the nonlinear saturation regime, allowing gluon-saturation effects to be studied through gluon Wigner distributions \cite{Hagiwara:2016kam}. The elliptic gluon distribution, corresponding to the \(\cos 2\phi\) harmonic of the Wigner distribution, represents the leading azimuthally asymmetric contribution. This angular correlation probes the correlation between the transverse momentum of the produced dijet and the recoil momentum of the nucleon, thereby providing access to transverse momentum--position correlations of gluons inside hadrons. Related angular structures have also been studied in diffractive transverse-momentum-dependent gluon distributions at small $x$ \cite{Hatta:2024dtmd}. Such correlations exist for the Weizs\"{a}cker-Williams-type gluon Wigner distributions as well, and an additional power-suppressed correction generates a $\cos 4\phi$ modulation in the dijet cross section \cite{Dumitru:2016jku,Hagiwara:2021xkf}.  \\

A complementary way to access the gluonic structure of hadrons is through energy-flow operators. Diffractive dijet production provides access to angular correlations of the gluonic structure, but it involves identified final-state configurations. Energy-flow observables instead provide a calorimetric way of accessing analogous information directly from energy deposition. These operators measure the energy flowing into different directions on the sphere at null infinity. Correlations of multiple energy-flow operators define the energy-energy correlator (or EEC) \cite{PhysRevLett.41.1585,PhysRevD.19.2018}. Owing to the energy weighting and the inclusive sum over collinear splittings, the EEC is infrared and collinear safe \cite{Kinoshita:1962ur,Lee:1964is}. In the small-$x$ regime, related energy-flow observables, including the nucleon energy-energy correlator (NEEC), have also been proposed as probes of gluon saturation \cite{Liu:2023aqb}; see also Ref.~\cite{Wang:2026efm}. These developments motivated the use of energy flow operator as a calorimetric probe of the angular structure generated by small-$x$ scattering. Furthermore, we investigate the conditions under which the complete angular intensity distribution can determine the underlying coherent diffractive amplitude itself. This formulation treats all even harmonics on an equal footing, makes the obstructions due to nodes and undetermined signs explicit, and exposes the autocorrelation ambiguity that arises once the harmonic-dependent phases are allowed.   \\

In the present work, we investigate to what extent energy-flow measurements in coherent diffractive dijet production can reconstruct the angular multipoles of the small-$x$ diffractive amplitude and thereby constrain the underlying dipole and Wigner harmonics. We first decompose the impact-parameter dependent dipole amplitude into its even angular harmonics and establish how these harmonics map onto the corresponding Wigner-distribution harmonics. Furthermore, we focus on the longitudinal photon channel, where, under the approximations specified below, the C-even dipole amplitude leads to a common phase for its angular harmonics. We then derive the angular structure of the diffractive scattering amplitude and show that each dipole harmonic maps directly onto an amplitude harmonic.\\ 

Moreover, we formulate the corresponding energy-flow measurement and relate its angular distribution to the squared scattering amplitude. In the node-free regime, we exploit the common-phase structure of the longitudinal C-even amplitude to show that taking the square root of the normalized energy-flow distribution recovers the normalized amplitude, whose angular multipoles can then be obtained through a direct Fourier projection. Subsequently, we extend the analysis to a more general complex-amplitude case and identify the conditions under which this reconstruction ceases to be unique. Finally, we conclude by discussing the main sources of corrections to the reconstruction arising from harmonic-dependent phases, C-odd exchange and effects beyond the leading-eikonal approximation \footnote{For clarity, throughout the paper we use $\phi'=\phi_P-\phi_\Delta$ for the relative azimuth between the dijet momentum and the recoil momentum, $\theta=2\phi'$, and $x_n\equiv A_{2n}/A_0$ for the normalized real amplitude multipoles. The fixed hard-kinematic label is denoted by $\Phi=\{z,P,\Delta,Q\}$. Unless otherwise stated, angular averages are $\langle f\rangle_\theta\equiv\int_0^{2\pi}d\theta\,f(\theta)/(2\pi)$.}.

\section{Dipole Wilson-line operator and symmetry decomposition}
For a quark moving eikonally through the target color field, the Wilson line in the fundamental representation is given by
\begin{equation}
    U({\mathbf{x}})={\cal P} \exp \left[ig \int dx^- A^+_a(x^-,{\bf x})t^a\right].
\end{equation}
In the eikonal approximation, the quark follows a straight lightlike path. Its transverse coordinates are frozen during the short interaction, while the ordered exponential resums arbitrarily many soft scatterings from the target. A virtual photon can fluctuate into a $q{\bar q}$ pair, whose interaction with the target is represented by the color-singlet dipole operator, constructed from the quark and anti-quark Wilson lines. For a color dipole, the operator is given by
\begin{equation}
    {\hat S}_Y(r,b)=\frac{1}{N_c}\mathrm{Tr}\left[U\left({ b}+\frac{{ r}}{2}\right)U^\dagger \left({ b}-\frac{{ r}}{2}\right)\right], 
\end{equation}
where $r$ is the dipole separation and $b$ is the impact parameter.
  The physical dipole S-matrix $S_Y$ is obtained by averaging the color-singlet dipole scattering operator over the ensemble of the target color-field configurations. At leading eikonal order, the target-averaged dipole operator contains both C-even pomeron and C-odd odderon contributions, corresponding to its real and imaginary parts respectively. In the present analysis, we restrict ourselves to the C-even sector, making the averaged dipole operator, $S_Y$, real. Furthermore, the dipole scattering amplitude can be written as,
  \begin{equation}
      T_Y(r,b)=1-S_Y(r,b),
  \end{equation}
  which is also constrained to be real. For a C-even unpolarized target, the angular dependence obeys two distinct symmetries, $T_Y(r,b,\phi)=T_Y(r,b,-\phi)$ and $T_Y(r,b,\phi+\pi)=T_Y(r,b,\phi)$. The first, a reflection symmetry of the unpolarized target ensemble, removes sine harmonics, while the second, equivalent to reversing the dipole orientation $r\rightarrow-r$, removes odd cosine harmonics. Thus, we can write
  \begin{equation}
      T_Y(r,b,\phi)=T_0(r,b)+2\sum_{n=1}^\infty T_{2n}(r,b)~ \cos (2n\phi), \label{q2}
  \end{equation}
where $T_0$ is the azimuthally averaged dipole amplitude and $T_{2n}$ characterize the successive even amplitude multipoles, given by 
\begin{eqnarray}
    T_0(r,b)&=&\int^{2\pi}_0 \frac{d\phi}{2\pi}T_Y(r,b,\phi),\\
    T_{2n}(r,b)&=&\int^{2\pi}_0 \frac{d\phi}{2\pi}T_Y(r,b,\phi)~ \cos(2n\phi), ~~~~n\geq 1.
\end{eqnarray}

The Fourier transform of the dipole amplitude $T_Y$ with respect to the dipole separation, together with the transverse differential operator appearing below, gives the dipole-type Wigner distribution at small-$x$. The dipole gluon Wigner distribution schematically can be written as,
\begin{eqnarray}
    W_Y({ k},{ b})={\cal N}_W\int d^2 re^{-i{ k}.{ r}}{\cal D}_{r,b} T_Y({ r,b}), \label{q1}
\end{eqnarray}
where $k$ is the transverse momentum conjugate to the dipole separation $r$ and ${\cal D}_{r,b}=-\nabla^2_r +1/4\nabla^2_b$ is a rotationally invariant transverse differential operator. 
This differential operator commutes with the transverse orbital angular momentum operator $L_z$, which indicates the angular momentum diagonality at the operator level (see the Appendix A).\\

While the non-linear small-$x$ evolution of $T_Y$ may generate higher harmonics, the map from $T_Y$ to $W_Y$ is linear and rotationally covariant. Consequently, the angular harmonic labels are preserved under this map allowing the Wigner distribution to be decomposed into the corresponding $W_{2n}$ harmonics as,
\begin{eqnarray}
    W_Y(k,b,\phi_{kb})=W_0(k,b)+2\sum_{n\geq 1}W_{2n}(k,b)\cos[2n(\phi_k-\phi_b)],
\end{eqnarray}
where,
\begin{eqnarray}
    W_{2n}(k,b)&=&2\pi{\cal N}_W(-1)^n\int^\infty_0 rdr J_{2n}(kr)\widetilde{T}_{2n}(r,b), 
    %\widetilde{T}_{2n}&=&-\left[\partial^2_r+\frac{1}{r}\partial_r-\frac{4n^2}{r^2}\right]T_{2n}+\frac{1}{4}\left[\partial^2_b+\frac{1}{b}\partial_b-\frac{4n^2}{b^2}\right]T_{2n}
\end{eqnarray}
Here, $\widetilde{T}_{2n}$ denotes the differential transform of the dipole harmonic $T_{2n}$ (for explicit expression of $\widetilde{T}_{2n}$ see Appendix A).  The isotropic term $W_0(k,b)$ denotes the azimuthally symmetric component, whereas the coefficient $W_2(k,b)$ characterizes the leading elliptic correlation. Thus, the rotationally covariant linear map preserves the angular-harmonic label.  \\

\section{Longitudinal coherent diffractive amplitude}
In this section, we connect the dipole amplitude, $T_Y$, whose harmonics encode the gluon structure, with experimentally accessible observables. We consider the Born $q{\bar q}$ impact factor for the longitudinal fluctuation of a virtual photon before the dipole crosses the target, while retaining the eikonal multiple scattering encoded by the Wilson lines, and write the coherent diffractive dijet process as
\begin{eqnarray}
    \gamma^*_L(q)+A(P_A)\rightarrow q(k_1)+{\bar q}(k_2)+A(P'_A).
\end{eqnarray}
 Here, we define the hard relative momentum $P$, conjugate to the dipole separation $r$ as, $P=(1-z){ k}_1-z{ k}_2$, and the recoil transferred to the target $\Delta$, conjugate to the impact parameter $b$, as ${\Delta}=-({k}_1+{k}_2)$, \cite{Dominguez:2011wm}. At the leading eikonal order, the dipole amplitude enters directly into the scattering amplitude as,
\cite{PhysRevD.100.034007},
\begin{equation}
    {\cal M}_L({ P},{ \Delta})={\cal C}_L \int d^2rd^2b ~ e^{-i{ P}.{ r}-i{ \Delta.b}} {\cal \psi}_L(z,r;Q)T_Y(r,b,\phi_r-\phi_b),
\end{equation}
Here, we consider a longitudinally polarized photon, where the corresponding photon wavefunction, $\psi_L(z,r;Q)$, is real and depends only on the dipole size, $r$, and is therefore a rotational scalar. Thus, it neither generates additional angular structures nor introduces a harmonic-dependent phase. A transverse photon carries explicit transverse indices and helicity phases, which can generate additional angular structure even before the target is considered. Thus, the longitudinal channel becomes the cleanest starting point, allowing the angular structures of the scattering amplitude to be directly mapped onto the harmonics of  dipole amplitude, $T_Y$. Since the dipole amplitude contains only the even harmonics, using Eq.\eqref{q2} the scattering amplitude can be decomposed as,
\begin{eqnarray}
    {\cal M}_L(\phi')=A_0+2\sum_{n\geq 1}A_{2n}\cos(2n\phi'), ~~~~ \phi'=(\phi_P-\phi_\Delta), \label{q14}
\end{eqnarray}
where,
\begin{eqnarray}
    A_{2n}(P,\Delta)={\cal C}_L \int^\infty_0 rdr\int^\infty_0 bdb ~ \psi_L(r) T_{2n}(r,b)J_{2n}(Pr)J_{2n}(\Delta b). \label{q3}
\end{eqnarray}
To see the explicit angular projection, we consider the $2n$-th harmonic of the dipole amplitude and expand the  Fourier exponential in angular Bessel modes (see Appendix B). The angular integrations then set the Fourier modes $m=-2n$ and $l=+2n$ for the angular integrations of $r$ and $b$ respectively.\\

Moreover, when the C-even dipole correlator and the real longitudinal impact factors are inserted, the radial integral associated with every harmonic becomes real. Thus, from the assumptions made above, we can write $A_{2n}={\cal C}_L\widetilde{A}_{2n}$ with $\widetilde{A}_{2n}\in \mathbb{R}$, where the same fixed process-dependent constant ${\cal C}_L$ multiplies every harmonic. We therefore have,
\begin{eqnarray}
    \frac{A_{2n}}{A_0}\in \mathbb{R}. \label{q4}
\end{eqnarray}
 This establishes the common-phase property of the scattering amplitude - after one removes the same overall process-dependent phase from all harmonics, the reduced coefficients are real. It is non-trivial to show that the Fourier-Bessel projection for a C-even eikonal dipole amplitude and a longitudinal photon with a real scalar coordinate space wave function, does not generate a relative phase between different even harmonics. Here, we work in the standard near-forward eikonal treatment of the longitudinal momentum transfer, where the longitudinal skewness $\xi$ satisfies the condition $\xi\ll1$. The size of the residual correction is model dependent, and no universal numerical power correction is assumed. The nonzero-skewness gluon GTMD constructions considered here, provide a natural framework for testing this approximation \cite{Chakrabarti:2025gtmd}. In this approximation, it is not required for the transverse recoil to vanish and therefore preserves the amplitude's sensitivity to the impact parameter.\\

\section{Energy-flow operator and Born-level tomography}
The energy-flow operator provides a calorimetric description by weighting the energy deposited in a given direction. The energy weighting of the operator makes it compatible with the inclusive treatments of collinear splittings, thus making it a natural starting point for the present analysis. At Born level, its action simplifies considerably, allowing the connection between the scattering amplitude and the energy-flow observable to be made explicit. The energy-flow operator in the direction $\hat n$ is the light-ray operator defined as \cite{Sveshnikov:1995vi,Tkachov:1995kk,Korchemsky:1999kt,Bauer:2008dt,Hofman:2008ar,Belitsky:2013xxa,Belitsky:2013bja,Kravchuk:2018htv},

\begin{eqnarray}
    {\hat{{\cal E}}}(\hat{n})=\lim_{R\rightarrow \infty}\int^\infty_{-\infty}dt~{\hat{n}_i}T^{0i}(t,R\hat{n}),
\end{eqnarray}
The definition in terms of the stress tensor is independent of the hadron species. The energy-flow operator acts diagonally on asymptotic states and measures the energy carried by the particles in a given direction \(\hat n\). For a general final state \(|X\rangle\), its action is given by
\begin{eqnarray}
    {\hat{{\cal E}}}(\hat{n})|X\rangle = \sum_{j\in X} E_j\delta^{(2)}\left(\Omega_{j}-\Omega_{\hat{n}}\right)|X\rangle.
\end{eqnarray}
At Born level, the measured final state for the coherent diffractive dijet is $|X\rangle=|q(k_1)\bar{q}(k_2)\rangle$, for which the energy-flow operator acts diagonally on the asymptotic state according to
\begin{eqnarray}
    {\hat{{\cal E}}}(\hat{n})|q\bar{q}\rangle=\left[E_1\delta^{(2)}\left(\Omega_{k_1}-\Omega_{\hat{n}}\right)+E_2\delta^{(2)}\left(\Omega_{k_2}-\Omega_{\hat{n}} \right)\right]|q\bar{q}\rangle.\label{q6}
\end{eqnarray}
To make the conditioning on the hard event explicit, we denote by $\Phi\equiv\{z,P,\Delta,Q\}$ the fixed hard kinematics and introduce ${\cal P}_{\mathrm{diff},\Phi}$, which projects onto coherent diffractive final states in the chosen kinematic bin. We then define
\begin{eqnarray}
    \frac{d\Sigma_E}{d\Phi\,d\Omega_{\hat n}}=\langle i|\hat{S}^{\dagger}{\cal \hat{E}}(\hat{n}){\cal P}_{\mathrm{diff},\Phi}\hat{S}|i\rangle.
\end{eqnarray}
Evaluating the energy-flow operator on the Born $q\bar q$ final state gives the energy deposited in the detector direction $\hat n$. At fixed $\Phi$, the detector direction is identified with the direction of the outgoing partons. Consequently, in the Born $q\bar q$ impact-factor sector the relative-azimuth dependence of the energy-flow distribution is inherited from the squared scattering amplitude and can be written as
\begin{eqnarray}
    \frac{d\Sigma_E}{d\Phi\,d\phi'}=\omega_E(\Phi)|{\cal M}_L(\Phi,\phi')|^2, \label{q9}
\end{eqnarray}
where $\omega_E(\Phi)$ denotes the sum of the partonic energy weights associated with the selected detector bin at fixed $\Phi$. Any kinematic Jacobian arising from the conversion of the detector solid angle to the relative-azimuth variable is included in $\omega_E(\Phi)$. At the Born $q\bar q$ level this factor is independent of $\phi'$ at a fixed $\Phi$, and hence it does not generate an additional target-sensitive harmonic. However, the Wilson-line amplitude still contains the full leading-eikonal multiple scattering of the dipole from the target. Thus, Born level refers only to the projectile impact factor and the absence of additional final-state radiation. At this order, the target-sensitive azimuthal dependence of the energy-flow observable is therefore identical to that of the differential dijet intensity, apart from the scalar energy weight. The operator definition of the energy-flow observable also allows the same framework to be extended to configurations with additional radiation. To isolate the angular dependence, we define $\theta\equiv 2\left(\phi_P-\phi_\Delta\right)$, so that $\cos(2n\phi')=\cos(n\theta)$, and write the normalized unprojected angular energy-flow distribution as
\begin{eqnarray}
    p_E(\theta;\Phi)=\frac{d\Sigma_E/(d\Phi\,d\theta)}{\left\langle d\Sigma_E/(d\Phi\,d\theta)\right\rangle_\theta}.
\end{eqnarray}
At fixed $\Phi$, $\omega_E$ is independent of the azimuth and therefore cancels in the ratio. For notational simplicity, we suppress the fixed label $\Phi$ below and write $p_E(\theta)$. The resulting distribution contains the angular information accessible through this energy-flow observable, since normalization removes the overall normalization. As no particular harmonic is selected, $p_E$ retains the complete angular dependence accessible through the energy-flow observable. Using the even harmonic structure and the common-phase property of the longitudinal scattering amplitude established in Eq.\eqref{q4}, we see that $p_E$ is even under the transformation, $\theta\rightarrow -\theta$, and can therefore be Fourier expanded as,
\begin{eqnarray}
    p_E(\theta)=1+2\sum^\infty_{N=1}c_N\cos(N\theta),
\end{eqnarray}
where $c_N=\langle p_E(\theta)\cos(N\theta)\rangle_{\theta}$ denotes the experimentally accessible angular moments of the normalized energy-flow distribution. Since we have $\theta=2\phi'$, the harmonic index $N$ in the $\theta$-space expansion corresponds directly to the amplitude label $2n$. Thus, the coefficients $c_N$ probes structures involving $A_{2n}$. These coefficients characterize the complete angular distribution, while the subsequent reconstruction relates them to the amplitude harmonics $A_{2n}/A_0$. However, the detector does not measure the Wigner multipole directly. Rather, it measures energy deposited by the final-state partons, with its angular dependence determined by the squared scattering amplitude. Thus, the energy-flow measurement provides indirect access to the angular structure of the Wigner distribution through its common dependence on the underlying $T_{2n}$, rather than measuring the Wigner multipoles directly.\\

\section{Exact square-root reconstruction in the real-multipole sector}
The intensity measurements do not, in general,  determine the relative phases between different harmonics of the scattering amplitude. The common-phase property established above provides a special case in which these relative phases are absent. In the Born longitudinal C-even sector,  every harmonic carries the same overall process dependent phase. Thus, we can write,
\begin{eqnarray}
    {\cal M}_L(\theta)={\cal C}_L F(\theta)~~~~~~ F(\theta)\in \mathbb{R},
\end{eqnarray}
where ${\cal C}_L$ contains the common overall phase. Since, this factor is common to all the harmonics, therefore cancels from the normalized angular observables. The normalized intensity determines the magnitude of the reduced amplitude,
\begin{eqnarray}
    \sqrt{p_E(\theta)}=\frac{|F(\theta)|}{\sqrt{\langle F^2\rangle_\theta}}.
\end{eqnarray}
If the reduced scattering amplitude $F(\theta)$ remains node-free within the chosen kinematic interval, the remaining ambiguity is a single global sign. Once the sign convention is fixed, one may replace $|F|$ by $F$ in the reconstruction. Thus, the inverse problem can be reduced to a linear Fourier projection, and therefore does not require the solution of a nonlinear phase-retrieval problem. Instead of expanding the intensity, which produces the quadratic harmonic mixing, we expand $\sqrt{p_E(\theta)}$ which is linear in the underlying real amplitude. The Fourier orthogonality allows each amplitude harmonic to be extracted independently through projection onto the corresponding cosine basis. We write
\begin{eqnarray}
    F(\theta)=A_0\left[1+2\sum_{n\geq 1}x_n\cos(n\theta)\right], ~~~~~~ x_n\equiv\frac{A_{2n}}{A_0},
\end{eqnarray}
so that $\langle F\rangle_\theta=A_0$ and the normalized multipoles are obtained directly from the measured square-root spectrum. Consequently, after fixing the global sign branch, the reconstructed harmonic ratios follow directly as,
\begin{eqnarray}
    \frac{A_{2n}}{A_0}=\frac{\langle\sqrt{p_E(\theta)}\cos(n\theta)\rangle}{\langle\sqrt{p_E(\theta)}\rangle}. \label{q8}
\end{eqnarray}
This constitutes the central result of the present work. The nontrivial part is not the square-root reconstruction itself, but the common-phase property established above, which reduces the generally non-unique intensity-to-amplitude inverse problem into a real sign problem. Within the Born longitudinal C-even sector and a node-free domain, the normalized Fourier moments of the square root of the measured energy-flow distribution reconstruct the normalized amplitude multipoles exactly, $A_{2n}/A_0$, through a direct Fourier projection of the measured energy-flow spectrum.\\ 

\section{Complex amplitudes and harmonic mixing}
The result obtained in Eq. \eqref{q8} relies on the common-phase property of the Born longitudinal C-even amplitude. In this section, we relax the constraint and consider the complex coherent amplitude. We write the normalized amplitude as a Laurent series,
\begin{eqnarray}
    \frac{{\cal M}_L}{A_0}=\sum^{\infty}_{m=-\infty}~ a_m e^{im\theta}.
\end{eqnarray}
Here, $a_m$ are the complex Fourier coefficients of the normalized scattering amplitude, with $a_0=1$ following from the normalization by $A_0$. Using Eq.\eqref{q9}, the measured energy-flow distribution becomes,
\begin{eqnarray}
    \frac{1}{\omega_E|A_0|^2}\frac{d\Sigma_E}{d\Phi\,d\theta}=\sum_{m,l}a_m a^*_le^{i(m-l)\theta}.
\end{eqnarray}
If we take $N=m-l$ as the harmonic difference, we can write the coefficient of $e^{iN\theta}$ as,
\begin{eqnarray}
    s_N=\sum^\infty_{l=-\infty}a_{l+N}a^*_l.\label{q7}
\end{eqnarray}
Thus, Eq. \eqref{q7} makes explicit the harmonic mixing inherent in an intensity measurement. It follows directly from the Wilson-line amplitude and the diagonal action of the energy-flow operator. Here, $s_N$ is the unnormalized observable harmonic of the measured intensity. Its zeroth harmonic $s_0$ is simply the total angular-averaged intensity.\\

The assumption of real coefficients is not required until this point.
Thus, the measured harmonic is the autocorrelation of the underlying amplitude spectrum. In general, it is not a single amplitude harmonic, but contains the interference of all pairs of amplitude modes separated by the harmonic distance $N$. The real-multipole result discussed in the previous section corresponds to the special case in which the square root of the normalized intensity restores the underlying amplitude. For generic complex coefficients, however, the autocorrelation coefficient $s_N$ do not uniquely determine the phases of $a_m$ as additional phase-sensitive information are required. This is the standard non-uniqueness of the phase retrival. In specific cases,  additional information such as analyticity, support constraints, positivity in a suitable representation, or external interference information can reduce the ambiguity. We do not impose any such conditions in the present construction. Thus, the loss of uniqueness is not a failure of harmonic analysis, but the standard non-injectivity of recovering a complex spectrum from its autocorrelation.\\

We now specialize to the real multipole sector as discussed in the previous sections. Since the reduced amplitude is real and contains only the cosine harmonics, its Laurent coefficients satisfy $a_n=a_{-n}=x_n$. Furthermore, we define Fourier coefficients of the normalized energy-flow distribution $c_N=s_N/s_0$, as
\begin{eqnarray}
    c_N=\frac{2x_N+\sum^{N-1}_{m=1}x_m x_{N-m}+2\sum_{m=1}^\infty x_m x_{N+m}}{1+2\sum^{\infty}_{m=1} x^2_m}. \label{q10}
\end{eqnarray}
The above expression establishes the exact relation between the measured angular anisotropies and the underlying amplitude multipoles. 
In general, a measured harmonic is not a direct measurement of the corresponding amplitude harmonics. Instead, it receives contributions from an infinite tower of interference terms.
\\

\section{Graded anisotropy expansion and square-root deconvolution}
In the previous section, we see infinite contributions to the measured harmonic $c_N$ coming from different amplitudes $x_n$. This infinite mixing can be simplified if one assumes the target anisotropies to exhibit a smooth hierarchical suppression with increasing harmonic order. As an organizing assumption we assign the grading $x_n={\cal O}(\epsilon^n)$. The graded construction is applicable when a finite set of measured moments, rather than the fully resolved angular distribution, is used for the reconstruction. It also separated the contributions to a given harmonic that arises from the lower order harmonics. It is therefore not an alternative to the exact full-distribution inversion, but a finite-moment approximation adapted to harmonic analyses. At harmonic order $N$, both linear and quadratic contributions in Eq.\eqref{q10} scale as ${\cal O}(\epsilon^N)$, whereas the first difference-mode term and the normalization correction coming from $s^{-1}_0$ first appear at order ${\cal O}(\epsilon^{N+2})$. Retaining these terms through grade ${\cal O}(\epsilon^N)$ gives,
\begin{eqnarray}
    c_N=L_N+{\cal O}(\epsilon^{N+2}), ~~~~~~ L_N\equiv 2x_N +\sum^{N-1}_{m=1}x_mx_{N-m}.\label{q13}
\end{eqnarray}
The resulting recursive structure can be expressed more conveniently through generating functions. We define the one sided amplitude generating function and the observable generating function respectively as,
\begin{eqnarray}
    X(t)=1+\sum_{n\geq1}x_nt^n, ~~~~~~ {\cal C}(t)=\sum_{n\geq 1}c_nt^n.
\end{eqnarray}
At harmonic order $N$, the product $X^2(t)$ generates precisely the combination $L_N$ appearing in Eq.\eqref{q13}, since the coefficient of $t^N$ is $2x_N+\sum_{m=1}^{N-1}x_mx_{N-m}$. This implies the graded relation $1+{\cal C}(t)=X^2(t)$, up to corrections of grade ${\cal O}(\epsilon^{N+2})$ in the $N$-th harmonic. Thus, in the isotropic limit, $x_n\rightarrow0$, we take the branch that reduces smoothly to $X(t)=1$, and write
\begin{eqnarray}
    X(t)=\sqrt{1+{\cal C}(t)},\label{q12}
\end{eqnarray}
Thus, Eq. \eqref{q12} provides a graded deconvolution formula for the amplitude multipoles. Expanding the square root generates the deconvolved harmonics order by order, with the lower-harmonic products systematically subtracted from each higher harmonic (see Appendix C). \\

\section{Logarithmic harmonics}
A direct phenomenological consequence of this reconstruction arises for the fourth angular harmonic, if it can be extracted from the $\cos 4\phi$ energy-flow moment. As the measured energy-flow distribution is quadratic, the fourth harmonic will also receive contributions from the lower-order amplitudes as well. At the amplitude level, we can write
\begin{eqnarray}
    F(\theta)=A_0\left[1+2x_1\cos\theta+2x_2\cos2\theta+\ldots\right].
\end{eqnarray}
Squaring the amplitude generates a fourth angular harmonic even if the coefficient $x_2$ vanishes. At the graded level, the observed energy-flow coefficient is $c_2=2x_2+x_1^2+{\cal O}(\epsilon^4)$. The first is the genuine fourth amplitude multipole, while the second is a reducible contribution generated by the nonlinear mixing of the lower elliptic harmonic. This is exactly where the above developed reconstruction becomes useful. Within the graded approximation, the contributions coming from the lower harmonics can be subtracted through the deconvolution method. \\

While the square-root reconstruction provides the ordinary amplitude multipoles directly from the measured energy-flow distribution, we show a complementary transformation by taking the logarithm of  the normalized energy-flow distribution as,
\begin{eqnarray}
    \log p_E(\theta)=\kappa_0+2\sum_{n\geq 1}\kappa_n \cos n\theta.
\end{eqnarray}
The non-zero logarithm harmonics are obtained by the corresponding Fourier projections as,
\begin{eqnarray}
    \kappa_n=\langle\log p_E(\theta)\cos n\theta\rangle_\theta, ~~~~~~ n\geq 1.
\end{eqnarray}
Provided the chosen node-free branch also satisfies $F(\theta)/A_0>0$, we expand the angular dependence of the normalized amplitude as,
\begin{eqnarray}
    \log \frac{F(\theta)}{A_0}=2\sum_{n\geq 1}g_n \cos n\theta. \label{q17}
\end{eqnarray}
For $n\geq 1$, the two representations are related as, $\kappa_n=2g_n$. Thus, the logarithmic harmonics provide an alternative representation of the measured energy-flow distribution. Within the real common-phase sector, they are related to the Fourier coefficients of $\ln F(\theta)$ (see Appendix C). This representation also provides a convenient way to organize non-linear structures in the reconstructed amplitudes. The reconstruction derived above is formulated within the Born-level leading-eikonal approximation. Beyond this regime, additional contributions can modify the relation between the measured energy-flow harmonics and the underlying amplitude multipoles.  \\

\section{Phenomenological interpretation}
For phenomenological use, the node-free and hierarchical assumptions should be established bin by bin rather than imposed globally. A natural domain is the hard correlation regime, where $P$ and $Q$ are well above the non-perturbative regime, $z$ is away from the endpoints and small-$|t|=\Delta^2$ which is the characteristic of the coherent diffractive regime. Since, the diffractive dijet angular modulations depend appreciably on $P$, $\Delta$, $Q^2$, and small-$x$ evolution \cite{Mantysaari:2019csc,Boer:2021upt,Boer:2023gtmd}, the position of the first node and the degree of harmonic suppression can vary across the phase space and between targets. We therefore do not assign a universal boundary for either the first diffractive node or for the hierarchy $x_n={\cal O}(\epsilon^n)$. Instead, their validity can be checked directly from the measured angular distribution and from the independent model calculations. \\ 

Experimentally, coherent diffractive events requires an intact recoil hadron or nucleus, or an equivalent exclusivity selection, together with a rapidity gap. Therefore, forward tagging and veto detectors are important for such event selection at the EIC \cite{EICYellowReport:2021,Aschenauer:2025tag}. The variables $P$, $\Delta$, $Q^2$, and $z$ can then be binned event by event, while the detector acceptance and finite angular resolution are accounted for through the response matrix used to unfold $p_E(\theta)$. Moreover, since the reconstruction is formulated for the longitudinally polarized photon only, its contribution should be separated from the transverse contribution by exploiting their different $y$-dependence with data taken at different beam-energy configurations. Although the feasibility of Rosenbluth-type longitudinal separations at the EIC has been studied in detail for inclusive DIS \cite{JimenezLopez:2025fl}, a corresponding diffractive analysis will require its dedicated acceptance and systematic study. These detector effects do not alter the formal inversion, but they determine the precision with which $p_E(\theta)$ or the Fourier moments can be extracted.

\section{Robustness and controlled breakdown of real-multipole reconstruction}
Although Eq.\eqref{q4} establishes the common phase for the harmonics within the Born-level leading-eikonal description of the coherent longitudinal diffraction, it should not be promoted to an all-order theorem. Once the underlying assumptions are relaxed, relative phases between angular multipoles can appear. Below we discuss the leading mechanisms that can lead to this breakdown.

\begin{itemize}
\item Diffractive nodes and sign ambiguity: Even if the condition $A_{2n}={\cal C}_L\widetilde{A}_{2n}$ holds exactly, the normalized intensity determines,
\begin{eqnarray}
    \sqrt{p_E(\theta)}\propto|F(\theta)|,
\end{eqnarray}
and not the signed real amplitude $F(\theta)$. If $F$ crosses zero, a discrete sign ambiguity survives. Although Fourier transforms can generate diffractive dips and nodes in appropriate kinematic bins, the relevant node here is specifically a zero of the reduced angular amplitude $F(\theta)$ at fixed hard kinematics. It should therefore not be identified with a diffractive minimum in an angle integrated or $t$- integrated cross section. Thus, the exact reconstruction is valid either in a node-free angular or kinematic domain or piecewise between adjacent zeros. This sign ambiguity is logically distinct from the complex phase-retrieval problem that arises when the amplitude contains non-trivial complex phases.

\iffalse
{\color{Red}If $F$ crosses zero, a discrete sign ambiguity survives. The dipole Fourier transforms are known to generate diffractive dips and nodes in suitable kinematics. The relevant node here is a zero of the reduced angular amplitude $F(\theta)$ at fixed hard kinematics and should not automatically be identified with a diffractive minimum in an angle-integrated or $t$-integrated cross section. Thus, the exact reconstruction formula should be applied either in a node-free angular or kinematic domain, or piecewise between adjacent zeros. This sign issue is logically distinct from the complex phase-retrieval problem.}\\ \fi

    \item Real part and analyticity corrections: Even though the approximations listed in the above section remove the harmonic-dependent phases, small-$x$ dipole phenomenology often restores the real part of an otherwise absorptive diffractive amplitude through a dispersion-type correction \cite{Kowalski:2006hc}. This correction is commonly estimated as,
    \begin{eqnarray}
        \beta=\tan\frac{\pi\lambda}{2}, ~~~~~~ \lambda=\frac{\partial \ln A}{\partial \ln (1/x)},
    \end{eqnarray}
    so that the amplitude acquires an additional phase. If the same local-dispersion prescription is applied harmonic by harmonic, one would associate the logarithmic slope
\begin{eqnarray}
    \lambda\rightarrow\lambda_{2n}=\frac{\partial \ln |A_{2n}|}{\partial \ln (1/x)}.
\end{eqnarray}
Since no symmetry requires the different angular multipoles to have identical small-$x$ evolution, the relative phase generated by this prescription is schematically $\Delta\delta_{nm}\sim (\pi/2)(\lambda_{2n}-\lambda_{2m})$ when the slopes are small. Analyticity corrections can therefore generate harmonic-dependent phases as,
\begin{eqnarray}
A_{2n}=|A_{2n}|e^{i\delta_{2n}}, ~~~~~~ \delta_{2n}\neq \delta_{2m},
\end{eqnarray}
hence, the condition in \eqref{q4} no longer holds and the exact square-root inversion is lost.\\

\item Target-side sources of complex phases: Keeping the odderon component $O^{(-)}_{xy}$ makes the dipole matrix element intrinsically complex already at the operator level. Loop absorptive parts, real or virtual radiation and spin-dependent or sub-eikonal operators can introduce additional sources of relative phases. We therefore restrict the real-multipole and common phase contruction to the domain in which the required phase structure is established. Beyond that domain, the NLO analysis serves as a test of the reconstruction theorem rather than as an automatic extension of the leading order result.\\

\item Perturbative and final-state corrections: Beyond the Born-level leading-eikonal approximation, real and virtual radiation can modify the relation between the angular amplitude and the energy-flow distribution. The soft radiation and quantum evolution can affect azimuthal asymmetries and the transverse-momentum structure of diffractive gluon distributions \cite{Shao:2024dijet,Iancu:2026dtmd}. While the radiative corrections need not destroy the angular decomposition, but interference, recoil, or absorptive contributions may generate additional angular dependence or relative phases between different multipoles. Parametrically, these effects enter beyond the present leading result at ${\cal O}(\alpha_s)$ in the impact factor, with logarithmically enhanced regions requiring resummation rather than a fixed-order estimate. Consequently, the measured harmonics need not obey the Born-level autocorrelation in its present form, and the direct square-root reconstruction may receive corrections. A systematic treatment of these effects is beyond the scope of the present leading-eikonal analysis. \\

\end{itemize}

\section{Conclusion}
In this work, we started from the small-$x$ dipole amplitude and showed
that its angular harmonics can be mapped, through a rotationally diagonal transform, onto the corresponding
harmonics of the gluon Wigner distribution. We then showed that the same
dipole amplitude is mapped onto the scattering amplitude for coherent
diffractive dijet production in the Born $q\bar q$ impact-factor sector. For longitudinal, C-even
scattering in this sector, we established that the scattering amplitude
harmonics are real and share an overall common phase. Furthermore, we
showed that the normalized energy-flow distribution contains sufficient
information to reconstruct the normalized real scattering amplitude. In
a node-free region, its square root gives the normalized amplitude after
fixing the remaining global sign ambiguity by a chosen sign convention.\\

For a generic complex amplitude, the energy-flow harmonics are instead given by autocorrelations of the amplitude spectrum, and the phase information can not be reconstructed from the intensity alone. In the real-multipole sector, we showed that a hierarchical grading of the amplitude anisotropies allows the exact moment level mixing to be deconvolved systematically through the generating-function relation $X(t)=\sqrt{1+{\cal C}(t)}$. This gives an order by order recontruction of the amplitude multipoles and, in particular, clarifies the relation between the forth energy harmonic and the corresponding amplitude harmonic. The logarithmic representation
provides a complementary description in terms of cumulant-like amplitude
anisotropies.\\

The reconstruction remains conditional on the assumptions used in this article. We also discuss about the operational strategy for EIC implementation. It is therefore required to work in the bins of  $\Phi=\{z,P,\Delta,Q\}$, identify the coherent diffraction using forward exclusivity information, unfold the normalized angular distribution, and test the node-free condition in each bin before the inversion is applied. The nodes of the reduced scattering amplitude introduce sign ambiguities that cannot be resolved from the energy-flow intensity alone. Moreover, analyticity corrections, C-odd exchange, radiative effects, and sub-eikonal contributions can generate harmonic-dependent phases and modify the Born-level relation between the energy-flow distribution and the scattering amplitude. The reconstruction is therefore exact within the real, common-phase, node-free sector, while the grade deconvolution provides a complementary moment-level reconstruction within the same real multipole sector when the harmonic mixing is retained.

\bibliography{ref}

\vspace{5mm}

\appendix
\section{Angular diagonality of the differential operator}
We demonstrate explicitly that the differential operator appearing in the definition of the Wigner distribution does not mix different azimuthal harmonics. The Wigner distribution is defined as
\begin{eqnarray}
      W_Y({ k},{b})={\cal N}_W\int d^2 r~e^{-i{ k}.{ r}}{\cal D}_{r,b} T_Y({ r, b})
 \end{eqnarray}
 where,
\begin{eqnarray}
    {\cal D}_{r,b}=-\nabla^2_r +\frac{1}{4}\nabla^2_b.
\end{eqnarray}
To establish the angular diagonality of this operation, consider a single angular harmonic of the dipole amplitude,
\begin{eqnarray}
    f_{2n}(r,b)\cos (2n\phi).
\end{eqnarray}
where $\phi_r$ denotes the azimuthal angle of $r$. The transverse Laplacian in the polar coordinates in the $r$-plane is given by,
\begin{eqnarray}
    \nabla^2_r=\partial^2_r+\frac{1}{r}\partial_r+\frac{1}{r^2}\partial^2_{\phi_r},
\end{eqnarray}
The radial function $f_{2n}(r,b)$ has no dependence on $\phi_r$. Consequently, we can write
\begin{eqnarray}
    \nabla^2_r\left[f_{2n}(r,b)\cos (2n\phi)\right]=\left[\partial^2_r+\frac{1}{r}\partial_r-\frac{(2n)^2}{r^2}\right]f_{2n}(r,b)\cos (2n\phi).
\end{eqnarray}
Thus, the Laplacian changes the radial coefficient of the harmonic but leaves its angular dependence unchanged. In particular, there is no term proportional to $\cos(2m\phi_r)$ with $m\neq n$. The same argument applies to the impact-parameter Laplacian. Writing the Laplacian in polar coordinates in the $b$-plane gives,
\begin{eqnarray}
    \nabla^2_b\left[f_{2n}(r,b)\cos (2n\phi)\right]=\left[\partial^2_b+\frac{1}{b}\partial_b-\frac{(2n)^2}{b^2}\right]f_{2n}(r,b)\cos (2n\phi).
\end{eqnarray}
Therefore, for the harmonic under consideration, the corresponding angular dependence is acted upon diagonally. It follows that the full operator $\mathcal D_{r,b}$ acts on each harmonic independently as,
\begin{eqnarray}
    {\cal D}_{r,b} \left[T_{2n}(r,b)\cos(2n\phi)\right]=\widetilde{T}_{2n}(r,b)\cos(2n\phi), \label{q15}
\end{eqnarray}
where,
\begin{eqnarray}
    \widetilde{T}_{2n}(r,b)=-\left[\partial^2_r+\frac{1}{r}\partial_r-\frac{(2n)^2}{r^2}\right]T_{2n}+\frac{1}{4}\left[\partial^2_b+\frac{1}{b}\partial_b-\frac{(2n)^2}{b^2}\right]T_{2n}.
\end{eqnarray}
Equation \eqref{q15} makes the absence of angular mixing explicit. This operator maps the $2n$-th angular harmonic back into the same harmonic and modifies only its radial and impact-parameter dependence. This can further be expressed at the operator level. Restricted to functions of the relative azimuth $\phi=\phi_r-\phi_b$, the azimuthal angular-momentum operator is $L_z=-i\partial_\phi$. The transverse Laplacian contains the angular operator only through $\partial_\phi^2=-L_z^2$, while its remaining terms are independent of $\phi$. It therefore follows that
\begin{eqnarray}
[\mathcal D_{r,b},L_z]=0.
\end{eqnarray}
Hence $\mathcal D_{r,b}$ is diagonal in the angular-momentum basis. This establishes, at the operator level, that the differential operator entering the Wigner distribution cannot mix different azimuthal harmonics.\\

\section{Angular projection and the emergence of the Bessel moments}
We use the Fourier-Bessel identity,
\begin{eqnarray}
    e^{-ikr\cos(\phi_r-\phi_k)}=\sum^\infty_{m=-\infty}(-i)^m J_m(kr)e^{im(\phi_r-\phi_k)}.
\end{eqnarray}
Thus, the angular dependence of the target harmonic is transferred to the corresponding harmonic in momentum space, while its radial dependence is accompanied by the Bessel function $J_{2n}(kr)$. For a single target harmonic one needs,
\begin{eqnarray}
    I_{2n}&=&\int^{2\pi}_0 d\phi_r e^{-ikr\cos(\phi_r-\phi_k) }\cos\left[2n(\phi_r-\phi_b)\right]\\
    &=&2\pi(-1)^n J_{2n}(kr)\cos\left[2n(\phi_k-\phi_b)\right].
\end{eqnarray}
For completeness, the same result can be seen more directly by working with the complex harmonic. To obtain the angular projection, consider the \(2n\)-th harmonic in its complex form,
\begin{eqnarray}
    T_Y^{(2n)}(r,b,\phi_r,\phi_b)
=T_{2n}(r,b)e^{i2n(\phi_r-\phi_b)}. \label{q16}
\end{eqnarray}
The Fourier phases associated with the relative momentum $P$ and momentum transfer $\Delta$ are expanded as,
\begin{eqnarray}
    e^{-iPr\cos(\phi_r-\phi_P)}&=&\sum^\infty_{m=-\infty}(-i)^m J_m(Pr)e^{im(\phi_r-\phi_P)}\\
    e^{-i\Delta b\cos(\phi_b-\phi_\Delta)}&=&\sum^\infty_{l=-\infty}(-i)^l J_l(\Delta b)e^{il(\phi_b-\phi_\Delta)}
\end{eqnarray}
Substituting these expansions together with Eq. \eqref{q16}, the dependence on the two azimuthal angles becomes $e^{i(m+2n)\phi_r}e^{i(l-2n)\phi_b}$,
apart from phases involving $\phi_P$ and $\phi_\Delta$. The angular integrations therefore impose the selection rules
\begin{eqnarray}
\int_0^{2\pi}d\phi_r ~e^{i(m+2n)\phi_r}=2\pi\delta_{m,-2n}, ~~~~~~ \int_0^{2\pi}d\phi_be^{i(l-2n)\phi_b}=2\pi\delta_{l,2n}.
\end{eqnarray}
The $\phi_r$ integration selects the Fourier-Bessel mode $m=-2n$, whereas the $\phi_b$ integration selects $l=+2n$. Consequently, the \(2n\)-th target harmonic can only produce Bessel functions of order \(2n\). The Fourier-Bessel phase factors also cancel harmonic by harmonic, since $(-i)^{-2n}(-i)^{2n}=1$. Using 
$J_{-2n}(x)=(-1)^{2n}J_{2n}(x)=J_{2n}(x)$, the two Bessel factors can therefore be written with the same positive order,
\begin{eqnarray}
    J_{-2n}(Pr)J_{2n}(\Delta b)
=
J_{2n}(Pr)J_{2n}(\Delta b).
\end{eqnarray}
Thus, the remaining angular phases combine into $e^{i2n(\phi_P-\phi_\Delta)}$. Therefore, the scattering amplitude can be written as,
\begin{equation}
    {\cal M}_L({ P},{ \Delta})={\cal C}_L \int d^2rd^2b ~ e^{-i{ P}.{ r}-i{ \Delta.b}} {\cal \psi}_L(z,r;Q)T_Y(r,b,\phi_r-\phi_b),
\end{equation}
where ${\cal C}_L$ collects the overall normalization and any convention-dependent numerical factors or phases and
\begin{equation}
\psi_L(z,r;Q)= Qz(1-z)K_0(\epsilon r), ~~~~~~ \epsilon^2 = z(1-z)Q^2 + m_q^2.
\end{equation}
The scattering amplitude further decomposes as,
\begin{eqnarray}
    {\cal M}_L(\phi')=A_0+2\sum_{n\geq 1}A_{2n}\cos(2n\phi'), ~~~~ \phi'=(\phi_P-\phi_\Delta), 
\end{eqnarray}
Hence, after the angular integrations, the \(2n\)-th harmonic of the scattering amplitude takes the form
\begin{eqnarray}
 A_{2n}(P,\Delta)=\mathcal C_L\int_0^\infty r\,dr\int_0^\infty b\,db \psi_L(z,r;Q)T_{2n}(r,b)J_{2n}(Pr)J_{2n}(\Delta b),
\end{eqnarray}
 The above equation gives a definite $2n$-th angular harmonic in coordinate space which is mapped onto the same angular harmonic in momentum space, with its radial and impact-parameter dependence transformed by Bessel functions of the same order. Thus, the $2n$-th angular harmonic is mapped onto the corresponding pair of Bessel moments.\\

\section{Explicit harmonic deconvolution}
Starting from the exact mixing relation already derived in Eq. \eqref{q10}, the first two coefficients can be written as,
\begin{eqnarray}
c_1=\frac{2x_1+2x_1x_2+2x_2x_3+\cdots}{1+2x_1^2+2x_2^2+\cdots},
\end{eqnarray}
and 
\begin{eqnarray}
    c_2=\frac{ 2x_2+x_1^2+2x_1x_3+2x_2x_4+\cdots}{1+2x_1^2+2x_2^2+\cdots}.
\end{eqnarray}
It immediately shows that even $c_2$ is not simply $2x_2$. We define $C_{2N}\equiv c_N$ for the measured energy-flow harmonics and restore the physical labels, $c_1\equiv C_2$ and $c_2\equiv C_4$. Expanding the exact relation, the \(C_4\) harmonic takes the form,
\begin{equation}
C_4=2\frac{A_4}{A_0}+\left(\frac{A_2}{A_0}\right)^2+2\frac{A_2A_6}{A^2_0}+\cdots.
\end{equation}
Thus, the familiar lower-harmonic-square contribution is only the first member of an infinite harmonic-mixing tower. It should not be identified directly with $W_2^2$, since $A_{2n}$ and $W_{2n}$ are different transforms of the same dipole harmonics.\\

Moving to the generating-function relation introduced in Eq.\eqref{q12}, provides a graded deconvolution rather than a globally exact identity. Expanding the square root order by order gives the amplitude multipoles in terms of the measured harmonics. We define the reconstructed amplitude harmonics as,
\begin{equation}
\widehat C_{2N}\equiv
2\frac{A_{2N}}{A_0}=2x_N.
\end{equation} 
The first few deconvolved multipoles are
\begin{equation}
\widehat C_2=C_2,
\end{equation}
\begin{equation}
\widehat C_4=C_4-\frac{1}{4}C_2^2, \label{q18}
\end{equation}
\begin{equation}
\widehat C_6=C_6-\frac{1}{2} C_2C_4+\frac{1}{8}C_2^3.
\end{equation}
More generally, we can write
\begin{equation}
x_N=
\sum_{p=1}^{N}
\binom{1/2}{p}
\sum_{\substack{n_1+\cdots+n_p=N\\n_i\ge1}}
c_{n_1}\cdots c_{n_p},
\end{equation}
which, together with $\widehat C_{2N}$, gives the general deconvolution of the amplitude multipoles. The Eq. \eqref{q18} subtracts the leading reducible contribution from the  elliptic harmonic and within the graded approximation reconstructs $2A_4/A_0$ at that order. Thus, the fourth energy-flow harmonic $c_4$ does not directly corresponds to the fourth amplitude harmonic. The exact autocorrelation formalism above provides the corresponding mixing structure for higher harmonics
and allow the amplitude multipoles to be extracted systematically.\\

This contribution should also be distinguished from the subleading-power
$\cos(4\phi)$ mechanism discussed in the Wigner-distribution
phenomenology literature. The two effects correspond to different
origins of a fourth harmonic and should not be identified with one
another.
\\

Furthermore, we write the logarithmic energy-flow harmonics corresponding directly to cumulant-like amplitude anisotropies under the same real-multipole assumptions in Eq.\eqref{q17}. Defining the logarithmic energy-flow cumulants by $K_{2N}\equiv \kappa_N,$ the low-order graded expansion gives, 
\begin{align}
K_2 &= C_2, \\
K_4 &= C_4 - \frac{1}{2}C_2^2, \\
K_6 &= C_6 - C_2 C_4 + \frac{1}{3}C_2^3.
\end{align}
These logarithmic quantities are conceptually distinct from the deconvolved $\widehat{C}_{2n}$. The latter reconstruct the ordinary amplitude multipole $A_{2n}$, whereas the logarithmic coefficients encode the corresponding irreducible structures of $\log F$ within the positive node-free branch.

\end{document}